%% file: main.tex
\documentclass{aa}

\input{style/settings}

\usepackage{graphicx}
\usepackage{newtxtext}
\usepackage[smallerops,cmbraces]{newtxmath}
\usepackage{xcolor}
\definecolor{xlinkcolor}{cmyk}{1,1,0,0}
\usepackage[
  bookmarks=true,         % show bookmarks bar?/ Changed March 22, 2019, for improved accessibility
  pdfnewwindow=true,      % links in new window
  colorlinks=true,        % false: boxed links; true: colored links
  linkcolor=xlinkcolor,   % color of internal links
  citecolor=xlinkcolor,   % color of links to bibliography
  filecolor=xlinkcolor,   % color of file links
  urlcolor=xlinkcolor,    % color of external links
  final=true,
]{hyperref}

\input{style/packages}
\input{style/definitions}

\usepackage{lipsum}
\definecolor{lightblue}{rgb}{0.1,0.5,0.89}

\begin{document}

\title{Simulating the LOcal Web (SLOW)}
\subtitle{VII. Intergalactic magnetic field models for multi-messenger applications }
\titlerunning{Intergalactic Magnetic Field Models} 

% \subtitle{I. Overviewing the $\kappa$-mechanism}

\author{
    Johannes Stoiber\inst{\ref{inst:usm},\ref{inst:origins}}\thanks{jstoiber@usm.lmu.de}
    \and
    Klaus Dolag\inst{\ref{inst:usm},\ref{inst:mpa}}
    \and
    Francesca Capel\inst{\ref{inst:mpp}}
    \and
    Benjamin Seidel\inst{\ref{inst:usm}}
    \and 
    Michael Kachelrieß\inst{\ref{inst:ntnu}}
    \and
    Ludwig M. B\"oss\inst{\ref{inst:chic}}
    \and 
    Jenny G. Sorce\inst{\ref{inst:lille}, \ref{inst:saclay}}
}
\authorrunning{J.\ Stoiber et al.}

\institute{
    Universitäts-Sternwarte, Fakultät für Physik, Ludwig-Maximilians-Universität München, Scheinerstr.\ 1, 81679 München, Germany\label{inst:usm}\\
    \email{jstoiber@usm.lmu.de}
    \and
    Excellence Cluster ORIGINS, Boltzmannstraße 2, 85748 Garching, Germany\label{inst:origins}
    \and
    Max-Planck-Institut für Astrophysik, Karl-Schwarzschild-Str.\ 1, 85748 Garching, Germany\label{inst:mpa}
    \and
    Max-Planck-Institut für Physik, Boltzmannstr. 8, 85748 Garching, Germany\label{inst:mpp}
    \and 
    Institutt for fysikk, NTNU, Trondheim, Norway \label{inst:ntnu}
    \and
    Department of Astronomy and Astrophysics, The University of Chicago, William Eckhart Research Center, 5640 S. Ellis Ave. Chicago, IL 60637 \label{inst:chic}
    \and 
    Univ. Lille, CNRS, Centrale Lille, UMR 9189 CRIStAL, F-59000 Lille, France \label{inst:lille}
    \and 
    Université Paris-Saclay, CNRS, Institut d'Astrophysique Spatiale, 91405, Orsay, France \label{inst:saclay}
}

\date{Received XX Month, 20XX / Accepted XX Month, 20XX}

\abstract
{ The propagation of ultra-high-energy cosmic rays (UHECRs) and ultra-high-energy gamma-rays remains an open question in astroparticle physics, with the intergalactic magnetic field (IGMF) playing a crucial role in deflecting charged particles and shaping electromagnetic cascade spectra. Characterizing the IGMF across cosmic large-scale structure is therefore essential for interpreting multi-messenger observations and constraining the magnetogenesis scenarios that seeded it.}
{ We aim to provide accurate IGMF models to the astroparticle physics community and test their properties and robustness.}
{ We analyze IGMF models derived from the constrained cosmological simulation SLOW alongside a set of rescaled magnetic field models. We further introduce a novel algorithm to determine an \enquote{ideal position} for galaxies lying below the constraining power of the initial conditions, enabling accurate line-of-sight magnetic field extraction toward relevant sources.}
{ The models span a wide range of filling factors and sample distinct regions of the electron density–magnetic field strength phase space in filaments, while converging in the cores of galaxy clusters; the simulated field from SLOW best reproduces the IGMF derived from the electromagnetic $\gamma$-ray cascade. Models extracted using the introduced \enquote{ideal position} yield improved accuracy and may benefit multi-messenger studies more broadly. The large-scale structure drift of simulated clusters exploited by the algorithm also offers a potential route to refining the simulation's constrained initial conditions.} 
{}
%{Future work should extend these models to astrophysical seeding scenarios and stochastic primordial fields, as motivated by inflationary and cosmological phase-transition models.} 


\maketitle
%
%-------------------------------------------------------------------

\section{Introduction}
\label{sec:introduction}

\begin{figure*}
    \centering
    \includegraphics[width=\textwidth]{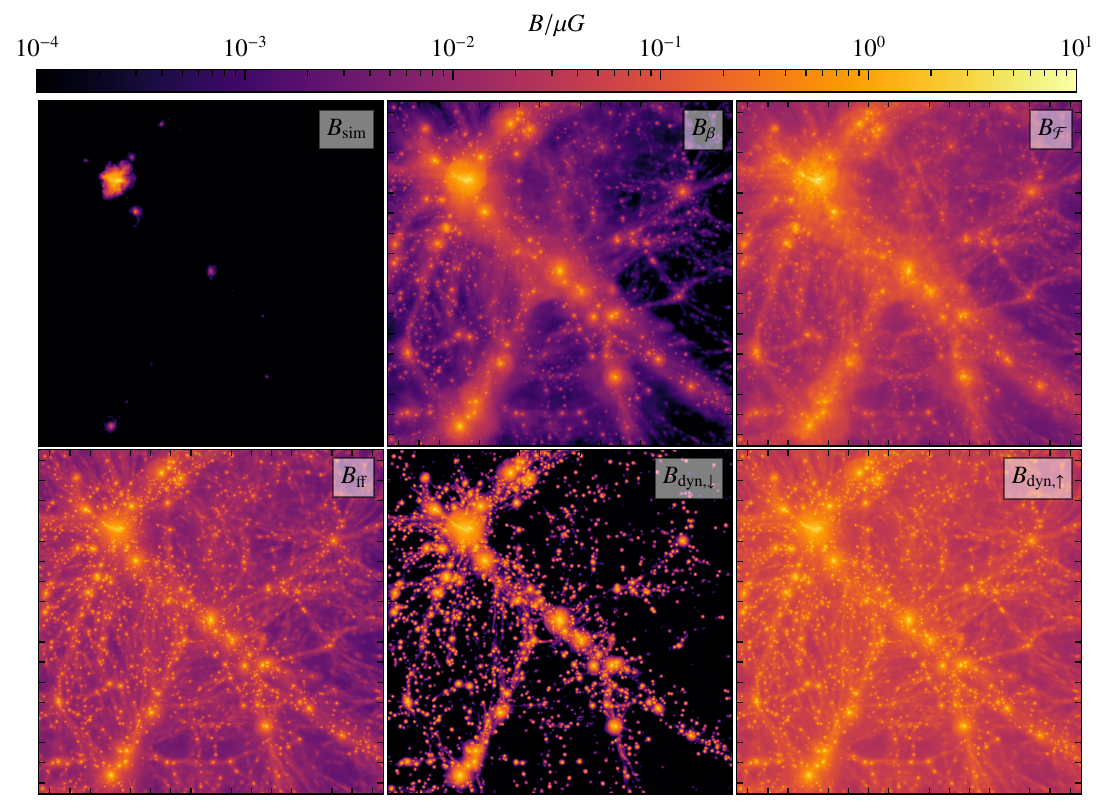}
    \caption{Cutout around a filament and the Coma cluster (top left corner) with a side-length of $\mathrm{d} \approx 34\ \mathrm{cMpc}/h$, showing the magnetic field strength for the six magnetic field models: Top row, left to right: $B_\mathrm{sim}$, $B_\beta$, and $B_\mathcal{F}$. Bottom row, left to right: $B_\mathrm{ff}$, $B_{\mathrm{dyn}, \downarrow}$, and $B_{\mathrm{dyn}, \uparrow}$. }
    \label{fig: maps}
\end{figure*}

The origin and nature of the intergalactic magnetic field (IGMF) remain some of the outstanding open questions in modern astrophysics. Its strength and topology affect the propagation of ultra-high-energy cosmic rays \citep[UHECR, $E \geq 10^{18}$ eV,][]{dolag+05a, hackstein+18} and TeV gamma rays \citep[e.g.][]{neronov&semikoz09, dolag+09-halos, dolag+11}, making multi-messenger observations a powerful probe of magnetogenesis in the large-scale structure of the Universe \citep{Durrer&Neronov2013}. Accurately modeling the IGMF along the line-of-sight to relevant astrophysical sources is therefore a critical ingredient in interpreting these observations.

Cosmological magnetohydrodynamic (MHD) simulations offer a physically motivated approach to constructing such models. Pioneering efforts by \citet{sigl+04} first estimated UHECR deflections through the simulated IGMF, finding large deflection angles. Later \citet{dolag+05a} introduced the important improvement of constrained initial conditions tied to the observed local large-scale structure, and consequently found deflections to be generally small outside of galaxy clusters. The benefit of using constrained simulations is in using these predictions in UHECR applications, for example, for backtracing real-world observed cosmic rays to their sources.

A promising field to probe the nature of magnetogenesis is the magnetic field in voids. The electromagnetic $\gamma$-ray cascade \citep{neronov&semikoz09} allows for deriving a lower limit on the IGMF strength along the line-of-sight to blazars \citep[e.g.][]{d_avezac+07, neronov&vovk10}. A TeV $\gamma$-ray emitted by a blazar will interact with the extragalactic background light (EBL) to produce an electron-positron pair, which travels in the same direction as the original $\gamma$-ray due to the high energy of the latter. Subsequently, the charged particles will inversely Compton scatter with the cosmic microwave background (CMB) and upscatter photons to GeV. Measuring a lower flux of GeV photons compared to the flux expected from the cascade theory points to the existence of a magnetic field, which deflects the charged particles off the line-of-sight \citep{d_avezac+07, neronov&semikoz09}. Most recent (conservative) results based on this effect are $B > 1.3 \times 10^{-17} \mathrm{G}$ \citep{blunier+25}.

Subsequent work used magnetic field models from constrained cosmological simulations \citep[\textsc{Coruscant},][]{dolag+05a} to study deflections of ultrahigh-energy cosmic rays \citep{dolag+05a}, and the electromagnetic pair-cascades \citep{dolag+11} to set a lower limit on the magnetic field strength and its volume filling factor $f(>B_\mathrm{th}) = \sum_i V(B_i > B_\mathrm{th})/V_\mathrm{tot}$, which is the fraction of the volume filled by a significantly strong magnetic field to cause a deflection off the line-of-sight. Additionally, the build-up of $\gamma$-ray halos due to the same effect \citep{dolag+09-halos} was studied. Other works utilizing cosmological MHD simulations include the systematic investigations of how different magnetogenesis scenarios affect the magnetic field in galaxy clusters \citep{donnert+09} or in general \citep{mtchedlidze+22}, or the anisotropy of UHECR arrival directions \citep{hackstein+18}\footnote{\citet{hackstein+18} use constrained simulations based on constraints gained with a method similar to the SLOW constraints.}. More recently, \citet{tjemsland+24} used the Chronos++ simulation suite \citep{vazza+22} to place lower limits on the strength of IGMF from $\gamma$-ray observations of the blazar 1ES 0229+200 to be $B > 5.1\times10^{-15} \mathrm{G}$, estimated the necessary filling factor to be $f \gtrsim 0.67$, and concluded magnetic field seeding based on the magnetization of the intergalactic medium by outflows of galaxies or AGN at $z \lesssim 6$ \citep[astrophysical seeding;][]{bertone+06, donnert+09, beck+13, vazza+25} as not sufficient to reach such high values. 

\citet{webar+25} recently claimed to have found significant evidence of an extended $\gamma$-ray halo around the TeV source Mrk\,501 and derived a root-mean-square magnetic field strength $B_\mathrm{rms} = 1.5 \times10^{-15} \mathrm{G}$ and a corresponding magnetic field correlation length $l_c = 10 \mathrm{kpc}$. 

Despite this progress, two recurring shortcomings limit the accuracy of existing magnetic field models for multi-messenger studies. First, many simulations employ unconstrained initial conditions, meaning that the simulated large-scale structure does not reproduce the observed local Universe \citep[e.g.][]{vazza+22, tjemsland+24}. Second, even simulations with constrained initial conditions are often too small to encompass the distances to many relevant sources without resorting to techniques such as mirroring the magnetic field profile at the box boundary \citep{dolag+11} or the concatenation of multiple simulation volumes \citep{tjemsland+24}, which inevitably reduces the fidelity of the resulting magnetic field profile along the line of sight. Additionally, interesting UHECR/$\gamma$-ray sources often are galaxy mass halos that are notoriously hard to replicate/cross-match even in constrained simulations, as the constraints are linearly reconstructed and the linear threshold lies around 3-4\,Mpc \citep{hernandezmartinez+24, seidel+24}. Another limitation shared by a broad class of UHECR source studies is the reliance on simplified IGMF representations, such as Gaussian random fields with a Kolmogorov turbulence spectrum \citep[e.g.][]{bister&farrar24, bourriche&capel+26}, in place of physical models extracted from simulations. 

In this work, we address these shortcomings by presenting magnetic field models extracted from SLOW \citep{dolag+23a}, a constrained cosmological MHD simulation of the local Universe at high resolution, complemented by rescaled magnetic field models \citep{boess+24}, and a novel algorithm for identifying an \enquote{ideal position} within SLOW for galaxies whose locations fall below the constraining power of the simulation's initial conditions by anchoring an \enquote{ideal position} to observed sources through the large-scale structure. 

We introduce the simulation suite in \cref{sec: simulation}, and discuss the properties of the resulting magnetic field models in \cref{sec: magnetic field}.  The algorithm to find an \enquote{ideal position} of unconstrained objects and the estimation of its uncertainty is described in \cref{sec: ideal pos} and \cref{sec: uncertainty}. We use the Monte-Carlo simulation code \texttt{ELMAG3.03} to analyze how the magnetic field models affect the $\gamma$-ray cascade spectrum in \cref{sec: elmag}. The resulting magnetic field models and their uncertainty estimates are presented in \cref{sec: results}. Finally, we discuss their robustness in terms of the $\gamma$-ray cascade and compare to previous work in \cref{sec: discuss}. In \cref{sec: conclusion} we conclude. %, specifically with the need for local universe simulations including additional magnetogenesis models \citep{donnert+09, mtchedlidze+22}. 

\section{Methods}

\subsection{Simulation}
\label{sec: simulation}

%The magnetic field models presented in this work are based on the constrained cosmological simulation SLOW (Simulating the LOcal Universe), which was already described in detail by \citet{dolag+23a}, \citet{hernandezmartinez+24}, \citet{boess+24}, \citet{seidel+24}, \citet{hernandezmartinez+25}, and \citet{boess+25}, so we will only give a short overview. 
The magnetic field models presented in this work are based on the constrained cosmological simulation SLOW (Simulating the LOcal Universe), which was already described in detail by previous work, which we briefly summarize below:

\subsubsection{Initial Conditions}

The initial conditions of a constrained cosmological simulation are not only imprinted with random Gaussian fluctuations but also include information about the real-world large-scale structure of the Local Universe. They come in three flavors: Constraints from galaxy number density, galaxy peculiar velocity measurements, or hybrid forms. A brief historical overview of different constrained Local Universe simulations can be found in \citet{dolag+23a}. The initial conditions for SLOW are based on the approach using galaxy peculiar velocities, which is described in detail by \citet{sorce18}. A summary of the most important steps in building these initial conditions, as well as a detailed description of the cross-identified structures and clusters, is provided by \citet{dolag+23a}, \citet{hernandezmartinez+24}, and \citet{seidel+24}. The observational dataset underlying the initial conditions is the Cosmicflows-2 catalog \citep[CF2,][]{tully+13}. The resulting initial conditions are based on the CLONE \citep[Constrained LOcal \& Nesting Environment,][]{sorce+21, sorce+24} simulations (in this case, realization number 8) that are also the basis of the HESTIA suite of simulations \citep{libeskind+20} and the simulations by \citet{hackstein+18} used a similar method to produce their constraints, among others. 

\subsubsection{The SLOW Simulation}

The SLOW suite of simulations includes several versions of various resolutions and the included subgrid physics \citep[see, ][]{dolag+23a, hernandezmartinez+24}. In this work, we use the SLOW-AGN$1536^3$ (also called SLOW-FP$1536^3$) simulation for positional data of clusters and galaxies that include a supermassive black hole (SMBH). More than 50 galaxy clusters have been cross-matched to real-world galaxy clusters so far \citep{hernandezmartinez+24, seidel+24}. Not only the large-scale structure \citep{dolag+23a}, but also the morphology and evolution of individual galaxy clusters, such as the bridge between A2667 and A3651 \citep{dietl+24} or the Fornax cluster \citep{reiprich+25}, were shown to match observations closely. Also investigated was the amount of turbulence in Coma, Virgo, and Perseus \citep{groth+26}. 

We use the SLOW-CR$3072^3$ simulation \citep{boess+24} to extract the IGMF. The latter simulation was run with the MHD solver \citep{dolag&stasyszyn09, bonafede+11} and the on-the-fly Fokker-Planck solver \textsc{CRESCENDO} \citep{boess+23} to model CR proton and electron evolution, but it is a pure non-radiative simulation with no star formation or SMBH. This simulation was already used to analyze the synchrotron emission from the simulated electron population and the simulated magnetic fields, as well as magnetic field models rescaled to the electron density, gas pressure, or turbulent pressure \citep{boess+24}. \citet{boess+25} also derived expectations for diffuse $\gamma$-ray emission from galaxy clusters and the cosmic web as modeled directly from the simulated CR energy density and spectra. Multi-messenger particles from galaxy clusters have also been modeled using this simulation and Monte Carlo simulations to propagate CRs \citep{hussain+25}. 

The two simulations are needed because SLOW-AGN$1536^3$ does not include a magnetic field but includes AGN that are needed to select clusters and galaxies that include an SMBH. SLOW-CR$3072^3$ in turn does not include AGN but includes a magnetic field. The positions of halos and subhalos, however, match very closely between the two simulations, as they use the same initial conditions and can easily be cross-matched. 

Both simulations have a volume of $(500 h^{-1} c \mathrm{Mpc})^3$. SLOW-AGN$1536^3$ consists of $2 \times 1536^3$ particles leading to a resolution of $M_\mathrm{gas} \approx 6.8\times10^{8} M_\odot$ and $M_\mathrm{DM} \approx 3.7\times10^{9} M_\odot$. SLOW-CR$3072^3$ consists of $2 \times 3072^3$ particles leading to a resolution of $M_\mathrm{gas} \approx 8.5\times10^{7} M_\odot$ and $M_\mathrm{DM} \approx 4.6\times10^{8} M_\odot$. The maximum spatial resolution at $z=0$ are $h_\mathrm{sml,min} = 5.97 \mathrm{\ kpc}$ and $h_\mathrm{sml,min} = 3.18 \mathrm{\ kpc}$, respectively. The gravitational softening of star and gas particles at $z=0$ are $\epsilon_\mathrm{stars} \approx 2 \mathrm{kpc}$, $\epsilon_\mathrm{gas} \approx 8 \mathrm{kpc}$ and $\epsilon_\mathrm{stars} \approx 1 \mathrm{kpc}$,  $\epsilon_\mathrm{gas} \approx 4 \mathrm{kpc}$, respectively. The simulations use Planck cosmology \citep{planckcollab14}: $\Omega_\mathrm{m} = 0.307$, $\Omega_\mathrm{baryon} = 0.048$, $\Omega_\Lambda = 0.692$, and $H_0 = 67.77 \mathrm{\ km\ s}^{-1}\mathrm{\ Mpc}^{-1}$.

\subsubsection{Simulation Code}

All simulations of the SLOW simulation suite were run with the \textsc{OpenGadget3} Code \citep{groth+23}, which is an improved implementation of the cosmological Tree-SPH code \textsc{Gadget2} \citep{springel05}. The physics subgrid model of SLOW-AGN$1536^3$ is mostly identical to the \textit{Magneticum} suite of simulations \citep{dolag+25}. For a detailed description of the subgrid model, see the same reference. The MHD implementation used for SLOW-CR$3072^3$ was presented by \citet{dolag&stasyszyn09}, and non-ideal MHD effects, such as magnetic diffusion and dissipation, were implemented by \citet{bonafede+11}. Divergence cleaning to enforce $\nabla\cdot\mathbf{B} = 0$ is implemented following \citet{tricco+16} as presented by \citet{steinwandel&price25}. Halos and subhalos are identified using \textsc{Subfind} \citep{springel+01}, which employs a standard friends-of-friends algorithm \citep{davis85} and is modified to include the baryonic component \citep{dolag+09-subfind}. 

\subsection{Magnetic field}
\label{sec: magnetic field}

SLOW-CR$3072^3$ was set up with a primordial magnetic field \cite[PMF,][]{kandus+11} as the seed magnetic field, as opposed to magnetic field seeding by the outflows of galaxies \citep[\enquote{astrophysical seeding}, e.g.,][]{donnert+09, beck+13, vazza+25, garg+25}. At the starting redshift of $z=120$, the magnetic field assigned to every gas particle is $\mathbf{B} = (10^{-14},0,0)$ G. This uniform magnetic field with constant strength and direction throughout the whole simulation box corresponds to an inflationary magnetogenesis scenario \citep{mukohyama+16} and is regularly employed in cosmological simulations \citep{dubois&teyssier08, marinacci+15, mtchedlidze+22, lehle+26}. Typically, inflationary magnetogenesis models are able to produce larger coherence lengths, as they are not limited by causality; however, they can also produce a stochastic distribution \citep[for a review,][]{kandus+11, Durrer&Neronov2013}. Another scenario for producing PMFs is cosmological phase transitions, which result in a stochastic distribution \citep{Durrer&Neronov2013}.  

The magnetic field predicted by SLOW was analyzed by \citet{boess+24} in absolute terms and with regard to reproducing diffuse synchrotron emission from cosmic web filaments, specifically around the Coma cluster. In the Coma cluster replica, the simulated magnetic field $B_\mathrm{sim}$ profile matches measurements using the rotation measure (RM) by \citet{bonafede+10} well up to a radius of $\sim 1\mathrm{Mpc}$ \citep{boess+24}. Further from the center, in the filaments, the magnetic field strength drops significantly, likely due to a lack of resolution to drive turbulent dynamo amplifications \citep{steinwandel+22}. To reconcile this mismatch, \citet{boess+24} adopted several models, rescaling the magnetic field strength based on the thermal gas pressure, turbulent pressure, or electron density.

The first model assumes a constant plasma-$\beta$ of $\beta = 50$ as the ratio between thermal pressure $P_\mathrm{th}$ and magnetic pressure as 

\begin{equation}
    B_\beta =  \sqrt{\frac{8\pi P_\mathrm{th}}{\beta}} \hspace{0.3cm} .
\end{equation}

The second model assumes the magnetic field as a fraction $\mathcal{F}$ of turbulent pressure, 

\begin{equation}
    B_\mathcal{F} = \mathcal{F} \sqrt{4\pi\rho v_\mathrm{turb}^2} \hspace{0.3cm} ,
\end{equation}

where we choose $\mathcal{F} = 1$, that is pressure equilibrium between turbulent and magnetic pressure. 

The following models scale with electron density. The third is derived from the ideal MHD flux-freezing argument, which states that magnetic field amplification occurs solely through collapse perpendicular to the field lines, following 

\begin{equation}
    B_\mathrm{ff} = \frac{B}{0.1\mu \mathrm{G}}\biggl(\frac{n_e}{10^{-4}\mathrm{cm}^{-3}}\biggr)^{2/3} \hspace{0.3cm} .
\end{equation}

The next models follow the assumption that the turbulent dynamo is the dominant amplification process, as found by \citet{steinwandel+24} in ultrahigh-resolution simulations of massive galaxy clusters. Therefore, the fourth model assumes that turbulent dynamo amplification works down to $n_e \sim 10^{-4} \mathrm{cm}^{-3}$, and below that limit, a fit was applied to capture the observations by \citet{carretti+23} with $B\sim 30 \mathrm{nG}$ in filaments of densities $n_e = 10^{-5} \mathrm{cm}^{-3}$ and extrapolated to $B\sim10^{-14} \mathrm{G}$ at $n_e = 10^{-6} \mathrm{cm}^{-3}$ as a stand-in for voids \citep[e.g.,][]{neronov&vovk10, tjemsland+24}, resulting in 

\begin{equation}
    B_{\mathrm{dyn},\downarrow}(n_e) =  \left\{ 
                            \begin{array}{ll}
                                    2.5\times 10^{-6} \bigl( \frac{n_e}{10^{-3} \mathrm{cm}^{-3}} \bigr)^{1/2} \hspace{0.05cm} [\mathrm{G}], \mathrm{if\ } \frac{n_e}{\mathrm{\ cm}^{-3}} > 10^{-4}      \\
                                    \sum\limits_{i=1}^{5} p_i \log_{10}(n_e)^{i-1}              \hspace{0.8cm} [\log_{10} \mathrm{G}], %\mathrm{if\ } \frac{n_e}{\mathrm{\ cm}^{-3}} < 10^{-4} 
                                    \mathrm{else}
                            \end{array}\right. \hspace{0.05cm} ,
\end{equation}

where $p \in [-16.38, -16.0, -8.07, -1.71, -0.13]$. The last model represents the upper limit of the pure saturated dynamo fitted to results by \citet{steinwandel+24}, scaling with electron density as 

\begin{equation}
    B_{\mathrm{dyn}, \uparrow}(n_e) = 2.5\times10^{-6} \biggl( \frac{n_e}{10^{-3}\mathrm{\ cm}^{-3}}\biggr)^{1/2} [\mathrm{G}] \hspace{0.3cm}.
\end{equation}

For more details, we refer to \citet{boess+24}. As these models are scaled to match the central magnetic field in the Coma cluster, $B_\beta$, $B_\mathcal{F}$, $B_\mathrm{ff}$, and $B_{\mathrm{dyn},\uparrow}$ overestimate the magnetic field in the voids significantly, but might still be valid in filaments \citep{boess+24}. \citet{alonso-lopez+26} found that the $B_\mathcal{F}$ model best predicts the rotation measure (RM) scatter in the Shapley Supercluster Core. 

Figure\,\ref{fig: maps} shows the magnetic field strength of the six magnetic field models $B_\mathrm{sim}$, $B_\beta$, $B_\mathcal{F}$, $B_\mathrm{ff}$, $B_{\mathrm{dyn}, \downarrow}$, and $B_{\mathrm{dyn}, \uparrow }$ (from top left to bottom right), in a cutout of $\sim 34$ c Mpc$/h$ near the the Coma cluster replica. There is almost no magnetic field above $0.1$ nG for $B_\mathrm{sim}$ except for the Coma cluster itself, and the filament is entirely below $0.1$ nG. In all other models, the filament has a field strength above $\sim 0.1 \mu$G. However, they differ in contrast: $B_{\mathrm{dyn},\downarrow}$ produces some magnetic field strength in the filament, but sharply drops below $0.1$nG in the adjacent regions, while $B_{\mathrm{dyn}, \uparrow}$ produces a magnetic field strength above $\sim 0.1 \mu$G almost everywhere. The remaining models range somewhere in between those two extremes. 

\begin{figure}
    \centering
    \includegraphics[width=\linewidth]{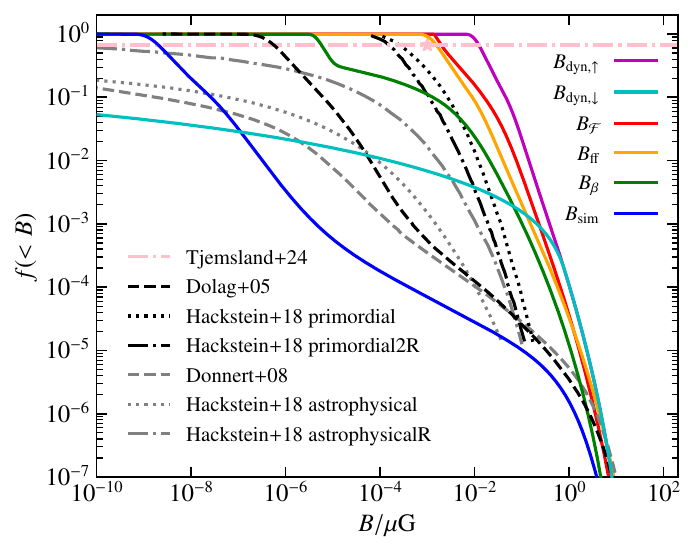}
    \caption{ Volume-weighted cumulative filling factors as a function of the corresponding threshold magnetic field strength for the six magnetic field models: $B_\mathrm{sim}$ (blue), $B_\mathrm{ff}$ (orange), $B_\beta$ (green), $B_\mathcal{F}$ (red), $B_{\mathrm{dyn}, \downarrow}$ (cyan), and $B_{\mathrm{dyn}, \uparrow}$ (magenta). The dashed black line represents the \textsc{Coruscant} Simulation \citep{dolag+05a}, and the dashed gray line its iteration including magnetic field seeding from galactic outflows instead of primordial magnetic fields \citep{donnert+09}. The dotted and dash-dotted black lines show simulations of primordial seeding by \citet{hackstein+18}, and the gray lines with the same styling show simulations of astrophysical seeding by the same author. The dashed-dotted horizontal pink line represents the lower limit of the filling factor for a tophat profile magnetic field at 1 nG estimated by \citet{tjemsland+24}.}
    \label{fig: filling}
\end{figure}

\begin{figure}
    \centering
    \includegraphics[width=\linewidth]{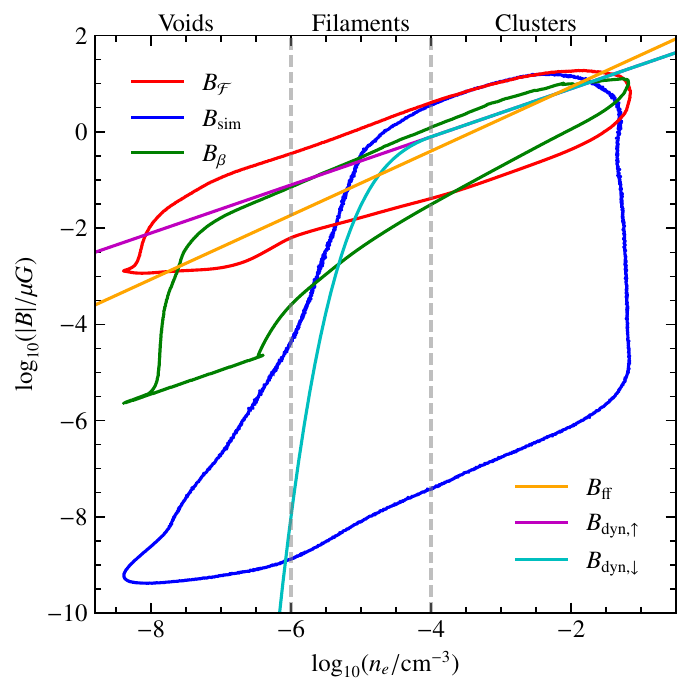}
    \caption{ Electron density-magnetic field strength phase space. For $B_\mathrm{sim}$, $B_\mathcal{F}$, and $B_\beta$ the $30\%$ contour of the mass-weighted 2D histogram is shown, while for $B_\mathrm{ff}$, $B_{\mathrm{dyn},\uparrow}$, and $B_{\mathrm{dyn},\downarrow}$ the model is a function of density anyway. See also \citet[][Fig.\,D.1.]{boess+24}.}
    \label{fig: phase space}
\end{figure}

The differences are more quantitatively depicted in \cref{fig: filling} that shows the volume-weighted cumulative filling factor $f(<B)$ as a function of the corresponding threshold magnetic field strength for the six magnetic field models. It describes what fraction of the volume of the simulation is permeated by at least the magnetic field strength on the x-axis. The simulated magnetic field $B_\mathrm{sim}$ (blue line) fills the least volume at a magnetic field strength down to $\sim 10^{-7} \mu$G, where $B_{\mathrm{dyn}, \downarrow}$ (cyan line) takes over as the least filling model. But, importantly, the simulated field fills the whole volume at a strength of $\sim 10^{-9} \mu$G, which is one order of magnitude below the seed PMF ($10^{-14}$ G)\footnote{This is the case, because magnetic fields not amplified by gravitational collapse or a turbulent dynamo are diluted due to the expansion of the Universe.}, while $B_{\mathrm{dyn}, \downarrow}$ does not. The model that reaches a filling factor of unity at the largest field strength ($\sim 10^{-2} \mu$G) is $B_{\mathrm{dyn}, \uparrow}$ (magenta line). $B_\mathrm{ff}$ (orange line) and $B_\mathcal{F}$ (red line) scale very similarly, both reaching a filling factor of unity at $\sim 10^{-3} \mu$G, while $B_\mathcal{F}$ fills a larger fraction at higher magnetic field strengths. $B_\beta$ (green line) scales similarly to the last four described molds at high $B$, but the corresponding filling factor flattens and only reaches unity at $2\times10^{-6} \mu$G after a sudden sharp increase. The shape of the filling factor of the simulated magnetic field has a similar shape to the filling factor in the \textsc{Coruscant} simulation \citep{dolag+05a} (black dashed line), but here the filling factor reaches unity at $\sim 2\times10^{-7} \mu$G. The difference stems from the higher seed PMF in \textsc{Coruscant} \citep[$2\times10^{-12}$ G to $10^{-11}$ G,][]{dolag+05a}. Simulations employing PMFs as magnetic seeds by \citet{hackstein+18} fill the whole volume at $\sim 10^{-4}$ regardless of whether they are seeded uniformly at $B_0 = 0.1$ nG (black dotted line) or statistically at $B_\mathrm{rms} = 1$ nG (black dash-dotted line). Finally, simulations with astrophysically seeded fields do not reach a filling factor of unity \citep[gray lines][]{donnert+09, hackstein+18}. Using a magnetic field model with repeating tophat profiles at magnetized pockets of $1$ nG, \citet{tjemsland+24} derived a filling factor of $f\gtrsim 0.67$, using the decline of GeV $\gamma$-rays in the spectra of blazars subject to the $\gamma$-ray cascade \citep{neronov&semikoz09}, consistent with early estimates by \citet{dolag+11}. The previous astrophysically seeded simulations do not reach such a filling factor, and at $1$ nG, only fulfilled by $B_\mathrm{ff}$, $B_\mathcal{F}$, and $B_{\mathrm{dyn}, \uparrow}$. 

Additionally, \cref{fig: phase space} shows the electron density-magnetic field strength phase space for all models. It shows that all models overlap and predict a similar magnetic field strength in the galaxy cluster regime ($n_e > 10^{-4}$), reaching a maximum of $\sim 1 \mu$G, which agrees well with observations \citep{bonafede+10}. However, in the regime of filaments ($10^{-6} < n_e/\mathrm{cm}^{-3} < 10^{-4}$), the average magnetic field strength of $B_\mathrm{sim}$ is significantly below $B_\mathcal{F}$, $B_\beta$, $B_\mathrm{ff}$, and $B_{\mathrm{dyn}, \uparrow}$. This was already pointed out by \citet{boess+24}. Note that the most populated region of $B_\mathrm{sim}$ (i.e., the mean magnetic field strength per density bin) is well below the other models in the filament regime \citep[see Fig.\,D.1. by][]{boess+24}. 

%In conclusion, the SLOW offers a broad range of magnetic field models suitable for studying the IGMF based on various assumptions of magnetic field amplification.  

\subsection{Constraining an \enquote{ideal} position within SLOW}
\label{sec: ideal pos}

In the study of individual sources of TeV $\gamma$-rays or UHECR, the magnetic field along the line-of-sight (magnetic field profile) to individual galaxy clusters or galaxies that host a blazar is of peak interest \citep[e.g.,][]{dolag+11}. So far, only simulated galaxy clusters have been cross-matched with real-world galaxy clusters in SLOW \citep{hernandezmartinez+24, seidel+24}. This is due to the fact that the initial conditions are based on a linear reconstruction of the initial velocity field and therefore, scales below the linear threshold of $\sim 3$ Mpc (corresponding to the most massive clusters $M_{500} > 10^{14} M_\odot$) are not directly constrained \citep{hernandezmartinez+24}. However, \citet{seidel+24} showed that it is possible to reach galaxy clusters/groups of masses $M_{500} > 0.2\times10^{14} M_\odot$ by looking for simulated counterparts based on the position of a nearby, more massive, already cross-matched galaxy cluster. In a first step, they identified the most massive galaxy cluster in a local supercluster (Shapley, Perseus-Pisces, Coma, Virgo, Centaurus, and Hercules). In a second step, they searched for the secondary member by adding the relative position vector of the observed position of the secondary member and the most massive member to the simulated position of the most massive member, yielding an \enquote{ideal} position \citep{seidel+24}. Relative to this position, the simulated counterpart was identified.

Our goal is to produce realistic magnetic field profiles along the line of sight to promising sources of UHECR and TeV gamma rays. \citet{sorce+24} note that the Hubble diagram (peculiar velocity as a function of distance) is well reproduced along the line of sight through galaxy clusters in the CLONE simulation, particularly for the most nearby clusters.

In order to produce these realistic magnetic field profiles along the line-of-sight, we expand on the above approach below by taking into account the nearest three galaxy clusters to the observed position of a source. In doing so, we include the following considerations: 

\begin{figure*}[ht]
    \centering
    \includegraphics[width=\linewidth]{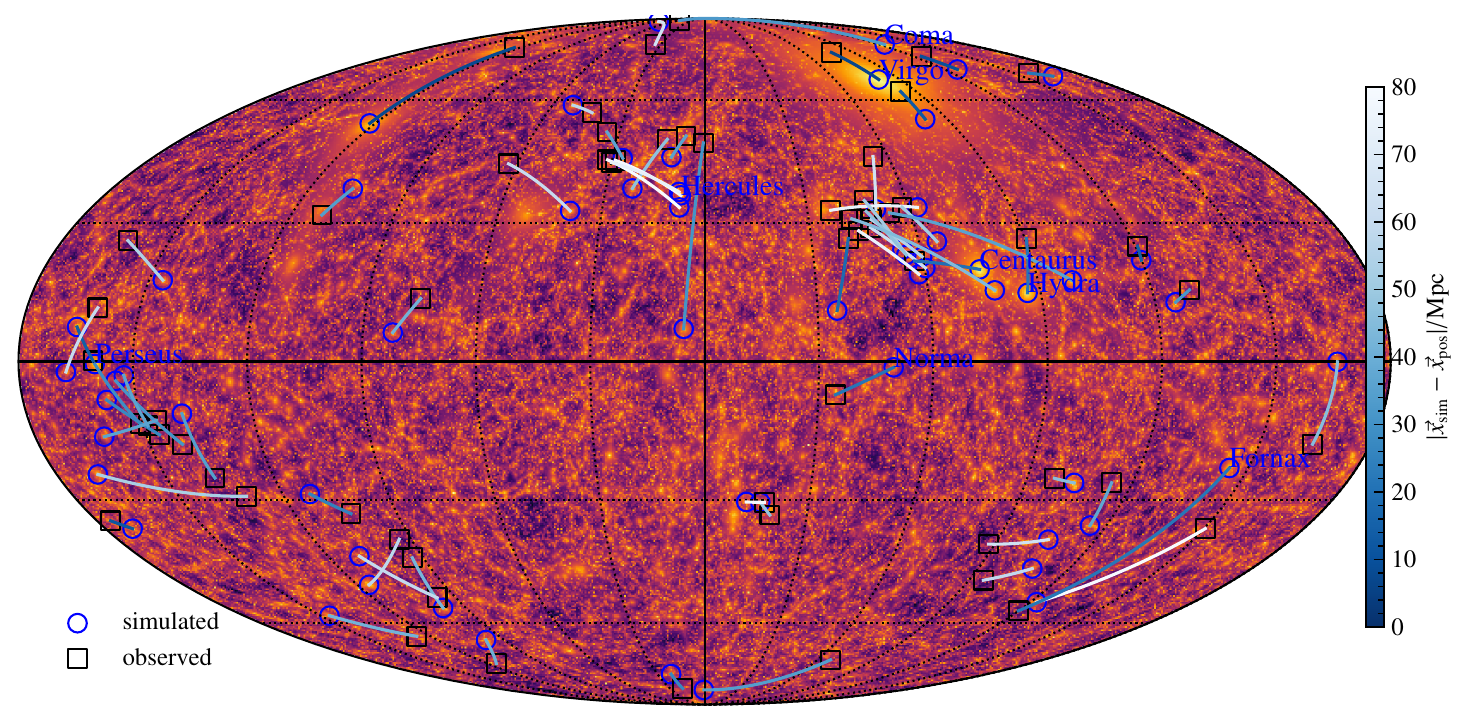}
    \caption{ Full-sky projection of the simulated (blue) and observed (orange) positions of prominent cross-matched galaxy clusters in SLOW \citep{hernandezmartinez+24, seidel+24}. The corresponding clusters are connected by lines colored by their 3D distance that follow sections of unit great circles. The background shows the X-ray surface brightness from sources out to a distance of 350 Mpc \citep{dolag+23a}. The coordinates are galactic with the Galactic Center at the origin, and positive longitude increasing to the left.}
    \label{fig: simobs}
\end{figure*}

\begin{enumerate}
    \item As we are interested in the properties along the whole line-of-sight, which foreground objects can influence, we select the three nearest galaxy clusters based on their position in the 2D plane of the sky (see below for more details). 
     \item The main criterion for the \enquote{ideal} position $\vec{x}_\mathrm{ideal}$ is that the distances between the observed position of a source $\vec{x}_\mathrm{observed}$ and the observed galaxy clusters $\vec{x}_i^o$ is approximately the same as the distance between the \enquote{ideal} position $\vec{x}_\mathrm{ideal}$ and the simulated galaxy clusters $\vec{x}_i^s$: $\delta x_i^o = |\vec{x}_\mathrm{observed} - \vec{x}_i^o | \approx |\vec{x}_\mathrm{ideal} - \vec{x}_i^s | = \delta x_i^s$. As this calculation is done on the 2D sky plane in galactic coordinates, the norm $| \cdot |$ represent the \emph{great circle distance}\footnote{$\theta_\mathrm{Great\ Circle} = \arccos \bigl( \cos{(\mathrm{lat}_1)} \times \cos{(\mathrm{lat}_2)} \times \cos{(\mathrm{long}_1 - \mathrm{long}_2)} + \sin{(\mathrm{lat}_1)} \times \sin{(\mathrm{lat}_2)} \bigr)$, where $\mathrm{lat}_i$ and $\mathrm{lon}_i$ are the latitudes and longitudes of the two positions \citep{mathworld+greatcircle}.} between the two angular positions. 
    \item However, the simulated large-scale structure could be offset from observations in a way that also changes the distance to the observer. Therefore, the ideal position for a galaxy in the simulation might also be further away than observed. For this reason, we use the three nearest galaxy clusters in 3D space to calculate the ideal distance (see below for more details). 
\end{enumerate}

Essentially, we want to use a large-scale drift of the simulated structure in reference to the observed structure to find an \enquote{ideal} position for a small-scale source within SLOW. This large-scale drift can be seen in \cref{fig: simobs}. In certain regions, it can be seen that simulated clusters (blue circles) have a systematic shift compared to observed clusters (black squares). For example, the Virgo cluster and clusters around it seem to shift collectively to the south. We speculate that this collective large-scale shift could be due to a yet unaccounted systematic in the construction of the initial conditions and could, in theory, be used to improve the accuracy of the initial conditions (Seidel et al., in prep.)\footnote{Other explanations include differences in simulation codes or the choice of $H_0$. However, we do not expect such large positional offsets purely from numerics \citep[e.g.,][]{sembolini+16}. The impact of the choice of $H_0$ will be tested in the future. }. 

The most straightforward way to calculate such an \enquote{ideal} position would be the following: 

\begin{enumerate}
    \item In a catalog of observed galaxy clusters, find the positions of the three closest galaxy clusters $\vec{x_i^o}$ ($i \in [1,3]$) to the observed position $\vec{x_\mathrm{observed}}$ of a given source.
    \item Calculate the relative position vector $\vec{x}_\mathrm{observed} - \vec{x}_i^o$ and add it to positions of the simulated galaxy clusters $\vec{x}_i^s$, resulting in \textit{shifted positions} $\vec{x}_i^\mathrm{shift}$.
    \item As it is highly unlikely that $\vec{x}_i^\mathrm{shift}$ are at same position for each $i\in[1,3]$, a consideration is building the centroid $\vec{x}_\mathrm{shift} = \frac{1}{3}\sum\limits_{i=1}^{3} \vec{x}_i^\mathrm{shift}$. 
\end{enumerate}

This centroid, however, is unlikely to satisfy the criterion $\delta x_i^o \approx \delta x_i^s$. Therefore, we apply what we call a \enquote{fuzzy triangulation}, that is, we calculate the parameter 

\begin{equation}\label{eq: Rparam}
    R = \sum_{i \in \mathrm{Nearest\ Clusters}} \frac{1}{\sqrt{2\pi}\sigma_i} \exp \biggl( -\frac{(\delta x_i^s - \delta x_i^o )^2}{2\sigma_i^2 } \biggr)  \mathrm{\ ,}
\end{equation}

in a grid around the shifted centroid $\vec{x}_\mathrm{shift}$. Because $R$ is $0$, when $\delta x_i^s = \delta x_i^o$, the position where $R$ is minimal is the \enquote{ideal} position, which needs to satisfy $\delta x_i^o \approx \delta x_i^s$. The choice of including a Gaussian, instead of simply calculating $R_\mathrm{strict} = \sum\limits_{i = 1}^3 \sqrt{( \delta x_i^s - \delta x_i^o )}^2$, leads to not strictly enforcing $\delta x_i^s = \delta x_i^o$ but allowing a \enquote{fuzzy shell} width. This helps with finding a global minimum instead of creating multiple local minima in the grid. We choose $\sigma_i = 10\times(\delta x_i^o)^{-1/2}$ as this width, which reduces the weight that the closest of the three nearest clusters has on the \enquote{ideal} position.

We note that performing such a \enquote{fuzzy triangulation} on three-dimensional positions can be used to identify simulated replicas of galaxy groups at masses even below $M_{500} \sim 0.2\times10^{14} M_\sun$. However, we do not claim that this is possible for all galaxies that might be sources of UHECR, such as Mrk501, Mrk421, or Mrk180 \citep{fang+14}. From the stellar mass-black hole mass relation in \textsc{Magneticum} \citep{dolag+25}, at a black hole mass of $\log{(M_\mathrm{BH}/M_\odot)} \sim 8.7$ \citep{ghisellini+10}, these galaxies have a stellar mass of $\log{(M_\star/M_\odot)} \approx 11.5$.

\begin{figure}
    \centering
    \includegraphics[width=\linewidth]{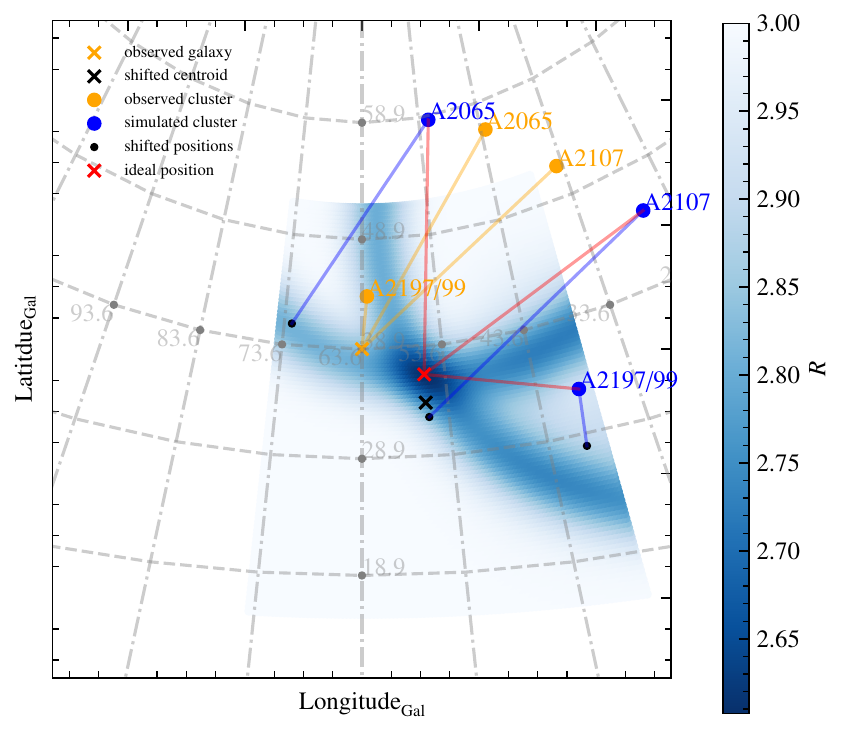}
    \caption{ Illustration of the grid search algorithm to minimize the differences between the observed distances (orange lines) and the distances expected from the simulation (red lines). The orange dots are the observed positions of the nearest three galaxy clusters to the observed position of a given source (orange cross, in this case, Mrk501). The blue dots are the simulated positions of the cross-matched galaxy clusters. The red cross shows the \enquote{ideal} position at which the parameter $R$ (blue-white color map) (see \cref{eq: Rparam}) is minimized, and therefore also the differences in the observed and simulated distances are minimal. The black cross and dots represent the \enquote{shifted} positions (see text).} 
    \label{fig: gridsearch}
\end{figure}

This process is illustrated in \cref{fig: gridsearch} for the position of Mrk501, which is an interesting source of $\gamma$ rays in the search for $\gamma$-ray halos around blazars caused by the existence of a significant IGMF \citep[e.g.,][]{webar+25}. 

In order to find the 3D position of this \enquote{ideal} position $\vec{x}_\mathrm{ideal}^{3D}$, we calculate the distance to the \enquote{shifted} centroid (see above), not using 2D coordinates, but 3D coordinates, and apply this distance to the angular position of the \enquote{ideal} position. 

Finally, each magnetic field model is calculated as the magnetic field along the line-of-sight towards this \enquote{ideal} $\vec{x}_\mathrm{ideal}^{3D}$. This is done by stepping along the line-of-sight in fractions of the local SPH smoothing length and evaluating the SPH function for the magnetic field at each position \citep{dolag+08-methods}:

\begin{equation}
    B(\vec{r}_i) = \sum\limits_j \frac{m_j}{\rho_j} B_j W(\vec{r}_i - \vec{r}_j, h_j) \mathrm{\ ,}
\end{equation}

where $\vec{r_i}$ are the positions along the line-of-sight and $\vec{r_j}$ are the positions of SPH particles, which smoothing kernels (of radius $h_j$) overlap with $\vec{r}_j$. $W(r,h)$ is the SPH smoothing kernel, where we used a $\mathrm{B}_2$-spline \citep{dolag+08-methods}. 

\subsection{Estimating Uncertainty}
\label{sec: uncertainty}

We estimate the uncertainty of these magnetic field profiles (along the line of sight) using three different methods and compare them to random lines-of-sight through the simulation box. All methods have in common that we calculate the magnetic field along multiple lines-of-sight towards positions offset from the \enquote{ideal} position, but still within a realistic spatial uncertainty scale $\lambda_\mathrm{D} = |\vec{x}_\mathrm{D, ideal} - \vec{x}_\mathrm{D, shift} |$, where $D$ is the dimension (either 3D or 2D), and $\vec{x}_\mathrm{shifted}^D$, is the \textit{shifted centroid} (see above). 

In the first method (\enquote{2D}), we choose 50 random positions inside a circle in the sky plane around the 2D \enquote{ideal} position with a maximum distance of $3\times\lambda_\mathrm{2D}$. The distances of these random positions are drawn from a Gaussian with a standard deviation of $\lambda_\mathrm{2D}$. 

In the second method (\enquote{3D}), we choose 50 random positions inside a sphere around the 3D \enquote{ideal} position with a maximum distance of $3\times\lambda_\mathrm{3D}$. The distances of these random positions are drawn from a Gaussian with a standard deviation of $\lambda_\mathrm{3D}$. 

In the third method (\enquote{subhalos}), we select galaxy mass halos in a sphere around the 3D \enquote{ideal} position with a radius of $3\times\lambda_\mathrm{3D}$. We select halos with a stellar mass of $10 \leq \log(M_\star/M_\odot) \leq 12$ and enforce that they host a supermassive black hole (SMBH), which is the basic requirement for hosting a blazar (an active galactic nucleus with a jet pointed towards the observer/earth). As the MHD simulation (SLOW-CR$3072^3$) does not contain black holes, we cross-match the subhalos of the MHD simulation with SLOW-AGN$1536^3$ that does contain black holes (See \cref{sec: simulation}). Again, we want to note that these are \textit{not} replicas of a galaxy constrained in SLOW, but are used to recreate a realistic environment for a UHECR or TeV gamma-ray source, which usually sits in the gaseous halo of a galaxy. The resulting number of suitable galaxies is of the same order as for the other two methods. 

%\begin{figure}
%    \centering
%    \includegraphics[trim={0 0 0 5cm},clip,width=\linewidth]{graphics/fig5.pdf}
%    \caption{ The expected position that is representative of the given blazar within the simulation (red cross, in this case Mrk501), together with the closest galaxies in the mass range $10 \leq \log(M_\star/M_\odot) \leq 12$ which host a supermassive black hole (blue circles). The white circles show the positions of the nearest cross-matched galaxy clusters. The coordinates are supergalactic, and the gas mass surface density colors the background.}
%    \label{fig: subhalos}
%\end{figure}

\subsection{Cascade spectrum}
\label{sec: elmag}

The last part of our methodology consists of employing the magnetic field profiles extracted using the method presented above in a Monte Carlo simulation of the electromagnetic $\gamma$-ray cascade using the code \texttt{ELMAG3.03} \citep{Kachelriess+2012, blytt+2020}. The magnetic field profile along the line-of-sight is gridded and read in as an input by the magnetic field module of \texttt{ELMAG3.03}. Employing such a magnetic field model within \texttt{ELMAG} was done by \citet{dolag+09-halos}, \citet{dolag+11}, and \citet{tjemsland+24}, among others. Our work improves on \citet{tjemsland+24} by using a constrained simulation and on \citet{dolag+09-halos, dolag+11} with the method to improve the pointing towards observed galaxies with the method outlined above.

\texttt{ELMAG3.03} simulates the process of the electromagnetic $\gamma$-ray cascade \citep[see, e.g.,][]{neronov&semikoz09, neronov&vovk10, Durrer&Neronov2013}: TeV $\gamma$-rays emitted by a blazar interact with the extragalactic background light (EBL) to produce an electron-positron pair in the inverse process of electron-positron annihilation. Subsequently, the charged particles inversely Compton scatter with and accelerate CMB photons to GeV energies. Due to the high energy of the initial $\gamma$-ray and relativistic beaming in the forward direction, both interactions do not significantly change the propagation direction of the involved particles. However, if there is a significant IGMF, the charged part of the cascade can be deflected, and the GeV $\gamma$-rays do not reach the observer. This can be seen as a decline of $\gamma$-rays in the GeV part of the spectrum.

We simulate this effect using our simulated magnetic field profile along the line of sight to the \textit{ideal} position of Mrk501 and a toy model for the source photon spectrum, not representative of an existing $\gamma$-ray source. The energy distribution is given by $\mathcal{F} \propto E^{-1.0}$ and a maximal energy $E_\mathrm{max} = 20\,\mathrm{TeV}$. This is a rather hard injection spectrum as observed for sources such as 1ES\,0229+200 \citep[e.g.,][]{tavecchio+10, dolag+11}. We also assume a jet opening angle of $\Theta_\mathrm{jet} = 6^\circ$ and $\Theta_\mathrm{offset} = 0^\circ$. The three-dimensional propagation mode is used to include our magnetic field model, binned to a one-dimensional grid. 

We produce a spectrum for each line-of-sight and each method to estimate the uncertainty of the line-of-sight, to calculate a measure of the uncertainty in a measure that conveys the actual impact of the different magnetic field strengths on observable properties. 

\begin{figure*}[ht]
    \centering
    \includegraphics[width=0.9\textwidth]{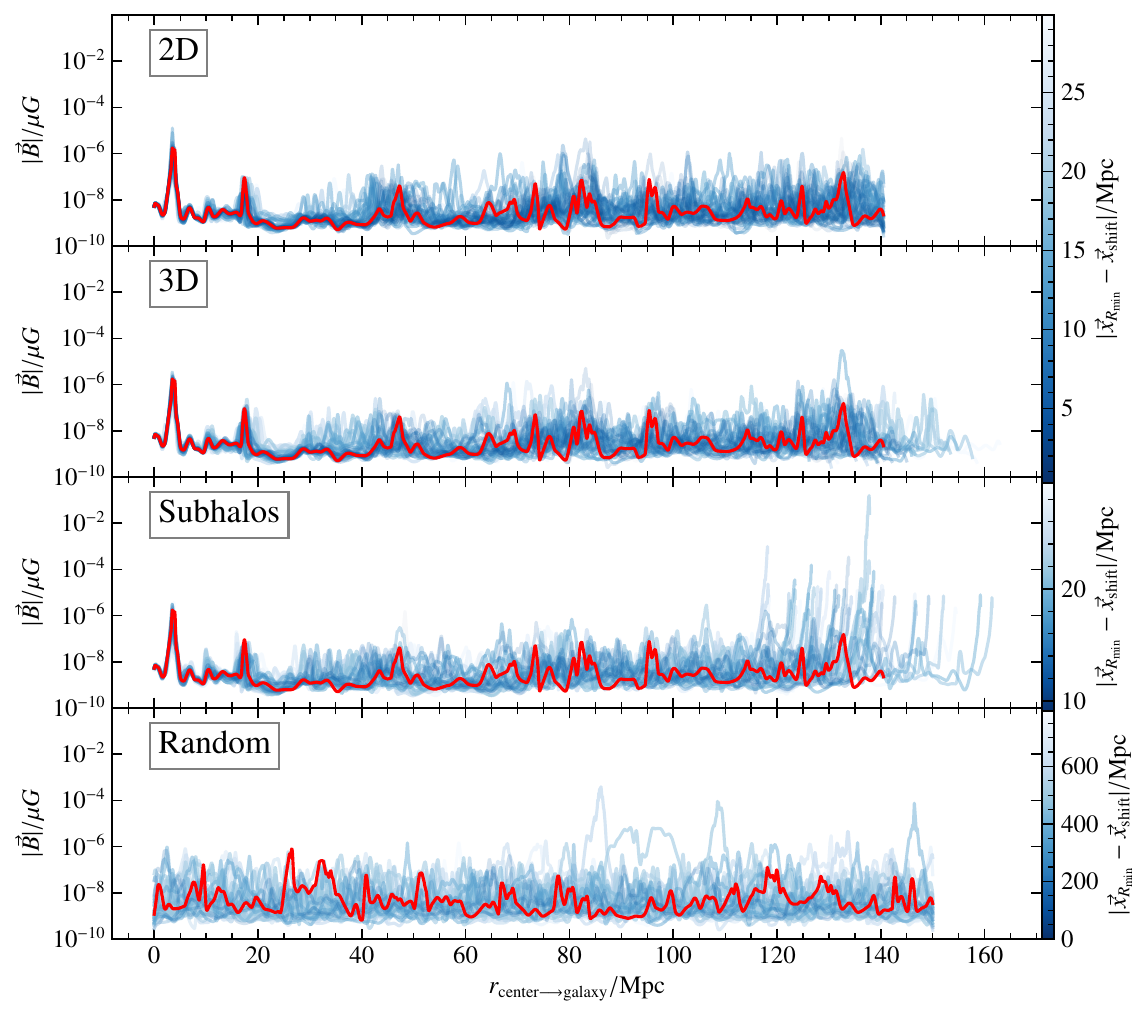}
    \caption{ The intergalactic magnetic field along the line-of-sight to the ideal position within the simulation (red, in panels one, two, and three). The bluish-to-white lines represent different methods to estimate the uncertainty of the magnetic field model (first to third panel: \enquote{2D shifting}, \enquote{3D shifting}, \enquote{suitable subhalos}, see text for details). The fourth panel shows a magnetic field model based on random lines-of-sight. The color represents the distance of the line-of-sight endpoint to the ideal position.} 
    \label{fig: losb}
\end{figure*}

\section{Results} 
\label{sec: results}

\subsection{Intergalactic magnetic field Profiles} 

We extract models of the extragalactic magnetic field from the constrained cosmological simulation SLOW by finding an \enquote{ideal} position within the simulation value for a given source of UHECR or TeV gamma-ray. We estimate the uncertainty of these magnetic field models by calculating the magnetic field along the line-of-sight to positions offset from the \enquote{ideal} position. Additionally, we use the Monte Carlo Code \texttt{ELMAG3.03} to estimate the propagation of this uncertainty when the magnetic field model is applied in a simulation of UHECR or TeV gamma-ray propagation. 

Figure\,\ref{fig: losb} presents the (simulated) magnetic field along the line-of-sight towards the \enquote{ideal} position of a given source (in this case Mrk501) in the SLOW simulation, and the magnetic field along the line-of-sight towards the offset positions as determined by the different methods to measure the uncertainty of the magnetic field (see \cref{sec: uncertainty}), in the first three panels. In the last panel, each line represents the magnetic field along the line-of-sight between a pair of random positions within the simulation box, but at the same distance as the distance to the \enquote{ideal} position from the observer. There are 50 such pairs. 

\begin{figure*}[ht]
    \centering
    \includegraphics[width=\textwidth]{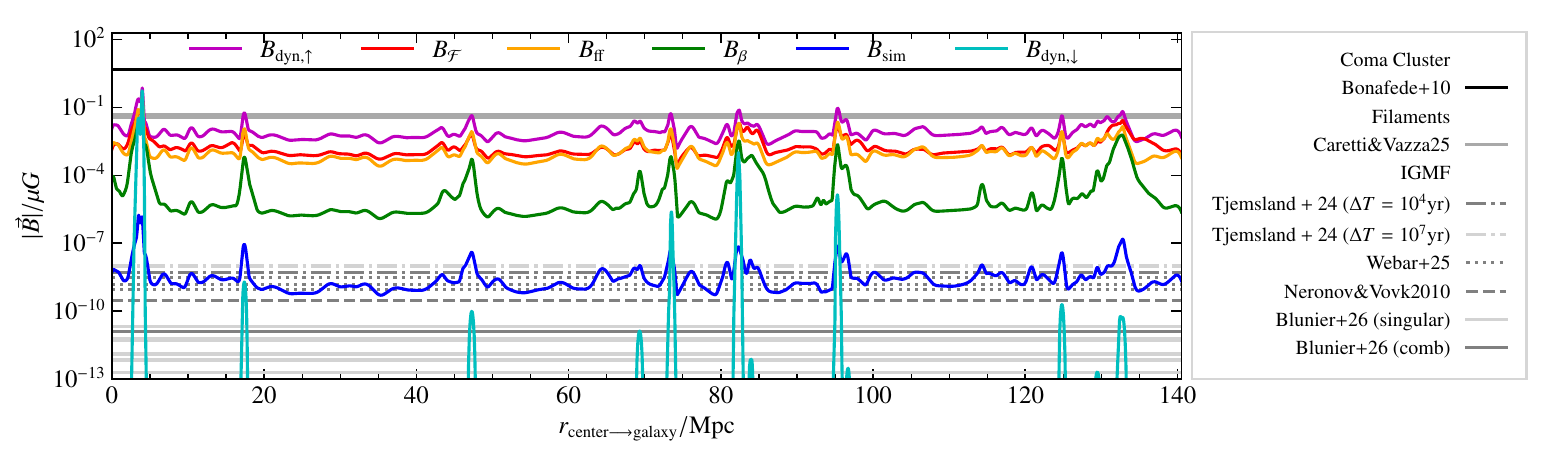}
    \caption{ The rescaled magnetic field $B_\mathrm{model}$ along the line-of-sight to the ideal position within the simulation. From top to bottom they are $B_{\mathrm{dyn}, \uparrow}$ (magenta), $B_\mathcal{F}$ (red), $B_\mathrm{ff}$ (orange), $B_\beta$ (green), $B_\mathrm{sim}$ (blue), and $B_{\mathrm{dyn}, \downarrow}$ (cyan). Also included are various observations of the magnetic field in different regimes of the Universe: The magnetic field in the center of the Coma cluster (black line, \cite{bonafede+10}), the magnetic field in filaments (dark gray line, \cite{carretti&vazza25}), and lower limits for the IGMF (\cite[dashdotdotted light gray line,][blazar activity of $\Delta T = 10^7 \mathrm{yr}$]{tjemsland+24}, \cite[dashdotdotted gray line,][blazar activity of $\Delta T = 10^4\mathrm{yr}$]{tjemsland+24}, \cite[dotted gray line,][]{webar+25}, \cite[dashed gray line,][]{neronov&vovk10}, \cite[solid light gray lines,][their results for singular blazars]{blunier+25}, and \cite[solid gray line,][their combined result]{blunier+25}. } 
    \label{fig: losb models}
\end{figure*}

The magnetic field along the line-of-sight to the \enquote{ideal} position (red line) takes on values roughly between $10^{-9} \mu\mathrm{G}$ to $10^{-6} \mu\mathrm{G}$ ($10^{-15}$ G to $10^{-12}$ G). We discuss in the next section how this compares to observations. The magnetic field towards the offset positions is in the same regime and has almost no difference in the first $10$ Mpc. The structure changes afterward, but is similar for each method. However, the last $\sim$ 10 Mpc vary significantly. In the case where the positions only vary in 2D, the magnetic field profiles are calculated up to the same distance, while in the \enquote{3D} and \enquote{subhalos} case, the distances vary.

Of particular note is the \enquote{subhalos} case, which stands apart from the other methods precisely because it incorporates a realistic environment as it inevitably ends within a cluster, while the other methods can end within a filament or void. This inclusion of realistic environmental structure is what drives a clear and distinguishing feature: an increase in the magnetic field in the last few Mpc, extending on average up to $10^{-4} \mu$G. This behavior is a direct consequence of modeling the environment realistically, rather than an artifact of position selection alone. By contrast, the purely random positions (bottom panel) yield a magnetic field strength in the same overall regime, but the scatter is larger in the first $10$ Mpc and also does not include a realistic environment in the last few Mpc.

To summarize, the \enquote{3D} and \enquote{subhalos} methods include a positional uncertainty not only in the angular position but also in the distance, while the \enquote{2D} methods only include angular uncertainty. These first three methods model the first few Mpc accurately because the position of the observer is clearly defined in the simulation. Additionally, the \enquote{subhalos} methods include a realistic environment in the last few Mpc. However, all magnetic field lines towards the ideal position create less scatter in the first few Mpc, compared to using random lines of sight. 

\subsection{Rescaled magnetic field profiles} 

So far, we have focused on the simulated magnetic field $B_\mathrm{sim}$ along the line of sight to a source. This model, directly obtained by solving the MHD equations, reproduces well the magnetic field strength in clusters \citep{boess+24}. However, in the density regime of filaments, it was shown to underpredict the magnetic field strength, resulting in effectively zero synchrotron emission \citep{boess+24}. Investigation of the magnetic field along a line-of-sight through a filament therefore needs to use different models, which we presented in \cref{sec: magnetic field}. 

Figure\,\ref{fig: losb models} presents the magnetic field along the line-of-sight to the ideal position of a source (here Mrk501) for all presented magnetic field models. Similar to \cref{fig: filling} and \cref{fig: phase space}, $B_{\mathrm{dyn}, \uparrow}$ results in the highest magnetic field followed by $B_\mathcal{F}$, $B_\mathrm{ff}$, $B_\beta$, and $B_\mathrm{sim}$. $B_{\mathrm{dyn}, \downarrow}$ is a special case, almost reaching $B_{\mathrm{dyn},\uparrow}$ in small overdensities \footnote{These are not yet galaxy clusters, which would reach $\gtrsim 1 \mu$G.}, but otherwise resulting in negligible magnetic field strengths. 

We compare the models to measurements of the magnetic field in the Coma cluster \citep{bonafede+10} and the average field strength in filaments \citep{carretti&vazza25}, both utilizing the rotation measure (RM), and several measurements of the IGMF \citep{neronov&vovk10, tjemsland+24, webar+25, blunier+25}, all utilizing the $\gamma$-ray cascade (see \cref{sec: elmag}). As this line-of-sight does not pass through a galaxy cluster, the magnetic field strength of the Coma cluster (solid black line) is never reached. The average field strength in filaments (solid gray line) is reached by $B_{\mathrm{dyn}, \uparrow}$ and $B_\mathrm{ff}$ in small overdensities, and once even by $B_{\mathrm{dyn}, \downarrow}$. The line-of-sight mostly passes through intergalactic space, which is best reproduced by the simulated magnetic field $B_\mathrm{sim}$, which is, over most of the distance, just slightly stronger than the measurement of the lower limit by \citet[][dashed gray line]{neronov&semikoz09}, \citet[][dash dotted gray lines]{tjemsland+24}, and \citet[][dotted gray lines, uncertainties are included]{webar+25}. Recent estimates by \citet{blunier+25} lie further below $B_\mathrm{sim}$, for both the strongest constraints by an individual blazar (light gray solid lines) and the combined constraints (dark gray solid line). However, these are the most conservative constraints as they assume a conservative blazar activity time scale of just $T\sim 16 \mathrm{yr}$ (the time span of Fermi/LAT data in their analysis). 

In conclusion, the available magnetic field models are able to reproduce all relevant regimes of the IGMF in the local Universe if combined correctly. 

\subsection{Cascade Spectrum}

\begin{figure}
    \centering
    \includegraphics[width=\linewidth]{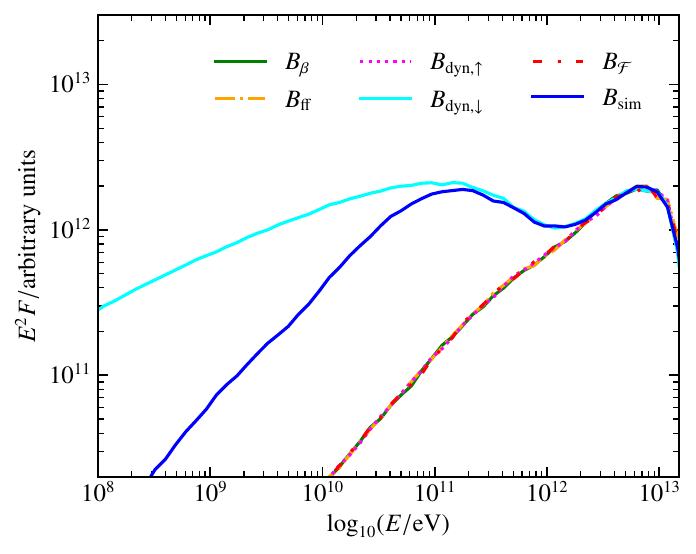}
    \caption{ Cascade spectrum for the different magnetic field models. Colors are the same as in \cref{fig: filling}, \cref{fig: phase space}, and \cref{fig: losb models}. } 
    \label{fig: cascade2}
\end{figure}

Finally, we want to investigate how the presented magnetic field affects the results of simulations of UHECR propagation or the gamma-ray cascade. In this work, we choose to simulate the electromagnetic $\gamma$-ray cascade using the Monte Carlo code \texttt{ELMAG3.03} as outlined in \cref{sec: elmag}. 

\begin{figure*}
   \centering
    \includegraphics[width=\textwidth]{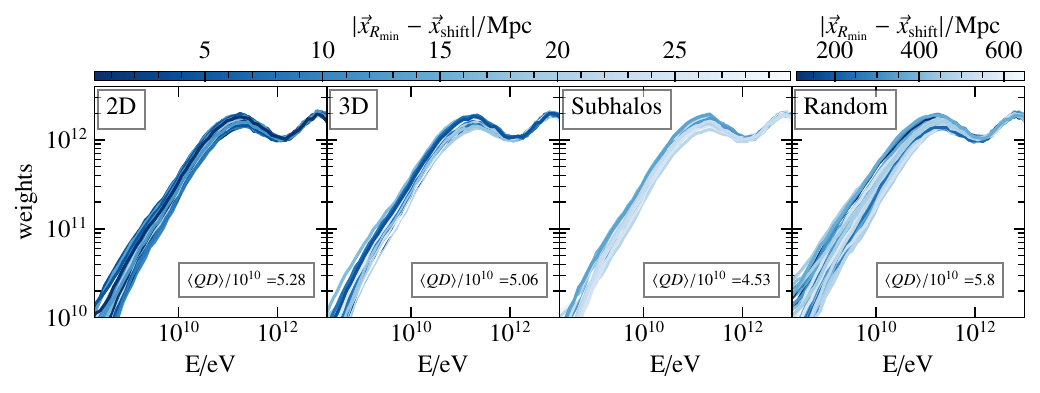}
    \caption{ Cascade spectrum for the different methods to estimate the uncertainty of the magnetic field as simulated with the Monte-Carlo Code \texttt{ELMAG 3.03}. First to third panel: \enquote{2D shifting}, \enquote{3D shifting}, \enquote{suitable subhalos}. Fourth panel: Cascade Spectrum for completely random lines-of-sight. The purely artificial gamma-ray source is at $z = 0.0412$ (173.4 Mpc) with a toy model of a source spectrum of $\mathcal{F} \propto E^{1.0}$ injected from 0.01 TeV to 20 TeV.}
    \label{fig: cascade}
\end{figure*}

We do not expect that $B_{\mathrm{dyn}, \uparrow}$, $B_\mathcal{F}$, $B_\mathrm{ff}$, and $B_\beta$ will give realistic results as they massively overpredict the magnetic field strength in voids. However, we include them for completeness. The spectra resulting from the $\gamma$-ray cascade being affected by the different magnetic field models are presented by \cref{fig: cascade2}. The spectrum at the high energy end ($E \gtrsim 10^{12}\,\mathrm{eV}$) is dominated by the primary (or injection) spectrum ($\mathcal{F}_0 \propto E^{-1}$, which is followed down to $E \sim 10^{10}\,\mathrm{eV}$ by $B_{\mathrm{dyn}, \uparrow}$, $B_\mathcal{F}$, $B_\mathrm{ff}$, and $B_\beta$, because secondary emission from the $\gamma$-ray cascade is entirely deflected away. $B_{\mathrm{dyn}, \downarrow}$ results in the most and $B_\mathrm{sim}$ in less but still significant secondary emission. This is the case, as a significant filling factor is needed to effectively deflect the electromagnetic part of the $\gamma$-ray cascade \citep{dolag+11}, which is not the case for void magnetic fields in $B_{\mathrm{dyn}, \downarrow}$ but is satisfied by $B_\mathrm{sim}$. Therefore, the results by \citet[][$f \geq 0.67$]{tjemsland+24} for the filling factor suggest that $B_\mathrm{sim}$ would be the better choice to model the IGMF for the $\gamma$-ray cascade. 

\section{Discussion}
\label{sec: discuss}

\subsection{Robustness of magnetic field models regarding the cascade spectrum} 

Finally, we discuss the robustness of the magnetic field models with respect to the cascade spectrum, that is, how much of the uncertainty is passed along to the cascade spectrum, and if it is less than when using random magnetic field models. 

Therefore, \cref{fig: cascade} presents the cascade spectra affected by the magnetic field models presented in \cref{fig: losb}, that is, the different methods of estimating the uncertainty of the magnetic field models and random lines-of-sight. For each method, we calculated the quartile distance\footnote{The quartile distance is the difference between the third quartile ($Q3$) and the first quartile ($Q1$) of a dataset: $QD = Q3 - Q1$. It measures the spread of the middle 50\% of the data \citep{mathworld+quartiledistance}. } per energy bin and built a mean quartile distance $\langle QD \rangle$ as an estimate of the uncertainty. The mean quartile distance is $\langle QD \rangle/10^{10} =$ 5.28, 5.06, and 4.53, for the \enquote{2D}, \enquote{3D}, \enquote{subhalos} method respectively, and 5.80 for random lines-of-sight. Using magnetic field models along the lines-of-sight to the \enquote{ideal position} results in a lower mean quartile distance than using random lines-of-sight. In fact, the lines-of-sight to the nearest suitable galaxies result in the smallest $\langle QD \rangle$, even though the average distance of their end points to the \enquote{ideal position} (color of all lines) is larger than the average distance of the end points gained in the \enquote{2D} and \enquote{3D} methods. We attribute this to the increase in the magnetic field in the last Mpc within the halos of galaxies, which is the only statistical difference from the latter methods. 

In conclusion, using magnetic field models extracted along the line-of-sight to a constrained \enquote{ideal position} from our Local Universe simulation reduces the uncertainty of the resulting $\gamma$-ray cascade spectrum. Specifically, using lines-of-sight with endpoints in realistic galaxy halos decreases the uncertainty further. 

\subsection{Comparison to previous work} 

Magnetic fields from cosmological simulations have been used in a multitude of works studying multi-messenger particles (i.e., $\gamma$-rays and UHECR). Here we list a selection. 

Pioneering work has been done by \citet{sigl+04}, who were the first to estimate the deflections of UHECRs through the IGMF as simulated by a cosmological simulation and using a source distribution drawn from the gas distribution of the same simulation within 50 Mpc. A crucial shortcoming was the use of an unconstrained cosmological simulation. This was improved upon by \citet[][\textsc{Coruscant}]{dolag+05a}, who used constrained initial conditions based on the IRAS 1.2-Jy all-sky redshift survey. They concluded that deflections of UHECRs ($E \geq 4\times10^{19}$ eV), accounting for structures within 110 Mpc, are small ($ < 1^\circ$) except in galaxy clusters.

Focusing on the magnetic field along the line-of-sight to singular objects, similar to what this study presents, \citet{dolag+09-halos} studied the effect of $\gamma$-ray halos building around point-like blazars caused by the electromagnetic pair-cascade \citep{neronov&semikoz09}. The same magnetic field models were used by \citet{dolag+11} to study the decline of the GeV $\gamma$-ray flux produced by the same cascade from TeV $\gamma$-rays and deflections caused by the magnetic field. A caveat of both studies was their limited simulation size, where the studied sources were located at distances $d_i \in [130,200]$ Mpc \citep{dolag+09-halos} or even $d \approx 591$ Mpc \citep{dolag+11}, while the size of the simulation was limited to $\approx 115$ Mpc.\footnote{One has to keep in mind, that even at a box size of 500 Mpc, only distances of up to 250 Mpc are covered, as the observer sits in the center of the simulation box. Additionally, the constraining power of the initial conditions does not cover the whole box but only reaches up to a distance of $\approx 200$ Mpc/h.} This required the mirroring of the line-of-sight magnetic field to extend the model, which potentially decreased the accuracy. 

More recent studies of UHECR propagation on the basis of cosmological simulations include \citet{hackstein+18}, who investigated the impact of different magnetogenesis scenarios and source distributions on the anisotropy of UHECR arrival directions. Models drawn from this simulation potentially reach large enough distances to include more of the relevant sources of UHECR or TeV gamma-rays as the simulation box size is $500$ Mpc/h (reaching distances of up to 250 Mpc/h). As already mentioned, their simulation use constraints were gained with a method similar to the SLOW constraints, but crucially were performed with the cosmological Eulerian MHD code \textsc{enzo} \citep{bryan+14}, instead of the Lagrangian SPH approach used for SLOW, offering the possibility to compare the two numerical methods. At a resolution of $(512)^3$ cells and dark matter particles, their resolution lies below SLOW-CR$3072^3$ at $(3072)^3$ gas and dark matter particles. Due to the use of adaptive mesh refinement, their resolution increases up to $\approx 31$ kpc/h per cell in the densest regions. The intrinsic adaptive increase of resolution in SPH pushed the maximum resolution of SLOW-CR$3072^3$ down to $h_\mathrm{sml, min} \approx 2.15$ kpc/h. 

Focusing on the magnetic field along the line-of-sight to singular objects, more recently, \citet{tjemsland+24} constrained the magnetogenesis scenarios and set lower limits for a space-filling primordial IGMF using the \textsc{elmag3.02} code, $\gamma$-ray data of 1ES 0229+200, and various magnetic field models extracted from the cosmological MHD simulation suite \enquote{Chronos++} \citep{vazza+22}. These are unconstrained simulations with a box size of $83.3$ Mpc. Therefore, the authors had to piece together multiple boxes to reach the distance to 1ES 0229+200 ($d \approx 591$ Mpc), and do not benefit from the constraints a constrained cosmological simulation provides. 

This list shows that many studies using magnetic field models from cosmological simulations either have the shortcoming of not using constrained initial conditions or not being large enough to cover distances to many relevant sources. Using magnetic field models extracted from SLOWCR-$3072^3$ can reconcile these issues. However, SLOW is also limited to a box size of $500$ Mpc/h (and therefore does not reach, for example, 1ES 0229+200) and does not yet provide the full range of possible magnetogenesis scenarios. The simulation by \citet{hackstein+18} also reconciles these shortcomings. We improve on it in resolution and by using a different numerical scheme, but we do not yet provide all their magnetogenesis scenarios. 

Crucially, we are the first to suggest an algorithm to find an \enquote{ideal position} for UHECR/TeV gamma-ray sources below the constraining scale of the initial condition and are therefore able to provide more accurate magnetic field models along the line-of-sight. 

Finally, studies that back-propagate UHECRs to find their sources often use very elaborate models of the galactic magnetic field \citep[e.g.,][]{unger&farrar2024}, but resort to employing a Gaussian random field with a Kolmogorov turbulence spectrum \citep[e.g.,][]{bourriche&capel+26} or an even simpler approach of Gaussian beam widening based on the same spectrum \citep[e.g.,][]{bister&farrar24}. The accuracy of such an investigation could potentially also be increased by employing magnetic field models as extracted from a cosmological simulation.

\section{Conclusion}
\label{sec: conclusion}

In this work, we present properties of the simulated magnetic field from the constrained cosmological simulation SLOW \citep{dolag+23a} and rescaled magnetic field models \citep{boess+24}. We present a novel algorithm to determine an \enquote{ideal position} for galaxies below the constraining power of the initial conditions within the simulation, in order to extract magnetic field models along the line-of-sight to relevant sources interesting for the study of UHECRs, ultra-high-energy gamma-rays, or the origin of the IGMF. 

We find that the presented magnetic field models cover a large range of magnetic field filling factors and cover different regions in the electron density-magnetic field strength phase in the density regime of filaments, but agree in the center of galaxy clusters. However, the IGMF agrees best with the lower limits derived from the gamma-ray cascade of GeV $\gamma$-rays from TeV $\gamma$-ray sources when using the simulated magnetic field. 

We present an algorithm to find an \enquote{ideal position} for galaxies in the SLOW simulation, whose positions are unconstrained by the initial conditions. The large-scale structure drift between observed and simulated galaxy clusters used in this algorithm can potentially also be used to improve the initial conditions. 

Finally, we show that the magnetic field models derived from this \enquote{ideal position} result in more accurate simulations of the $\gamma$-ray cascade spectrum and therefore might also benefit other multi-messenger studies employing a model for the IGMF. 

However, all current models are based on the uniform primordial magnetic field used in SLOW. Possible observables such as the proposed gamma-ray halo produced by the gamma-ray cascade are sensitive to the topology of the magnetic field lines. Therefore, it will be important to develop further magnetic field models based on astrophysical seeding scenarios \citep{donnert+09, vazza+25} and primordial magnetic field seeds following stochastic fields with and without helicity originating from inflation or cosmological phase transitions \citep{hackstein+18, mtchedlidze+22, schober+26}. 

\section*{Data Availability}

The magnetic field models are available on reasonable request to the first author and soon will be available online through the Cosmological Web Portal (\url{https://c2papcosmosim.uc.lrz.de}), where data products of the Magneticum Simulations \citep{dolag+25} are available, and data products of the SLOW Simulations are available to a selected user base (on reasonable request to K. Dolag). 

\begin{acknowledgements}
We thank Lucas Kimmig, Ildar Khabibullin, and Rhea-Silvia Remus for fruitful discussions. 

The hydrodynamical simulations were carried out at the Leibniz Supercomputer Center (LRZ) under the project pn68go (SLOW).
This research was supported by the Excellence Cluster ORIGINS, funded by the Deutsche Forschungsgemeinschaft under Germany's Excellence Strategy -- EXC-2094-390783311, and the ACME project, which has received funding from the European Union's Horizon Europe Research and Innovation program under Grant Agreement No 101131928.

%The following software was used for this work: Julia \citep{bezanson+14:julia}, GadgetIO.jl \citep{boess&valenzuela25}, GadgetUnits.jl \citep{boess25-Units}, SPHtoGrid.jl \citep{boess25-SPHtoGrid}, matplotlib \citep{hunter07}. 
The following software was used for this work: Julia \citep{bezanson+14:julia}, matplotlib \citep{hunter07}, GadgetIO.jl, GadgetUnits.jl, SPHtoGrid.jl (\url{https://github.com/LudwigBoess}). 
\end{acknowledgements}

\bibliographystyle{style/aa}
\bibliography{bib} 

\end{document}

%% file: style/settings.tex
\makeatletter

\renewcommand*\aa@pageof{, page \thepage{} of \pageref*{LastPage}}

\usepackage{natbib}
\bibpunct{(}{)}{;}{a}{}{,}%
\def\bibfont{\aa@bibliographyfont}%
\providecommand{\@LN}[2]{}

\makeatother

%% file: style/packages.tex
\usepackage{bm} % bold math
\usepackage{textgreek} % upright greek letters

\usepackage[capitalize]{cleveref} % easy referencing with \cref{fig:name} -> Fig.~X
\crefname{section}{Sect.}{Sects.}
\creflabelformat{equation}{#2#1#3} % equations without parentheses
\crefname{enumi}{item}{items} % items in lower case

\usepackage{siunitx} % proper unit typography
\DeclareSIUnit[number-unit-product = ]\percent{\char`\%} % no space before percent

\usepackage{csquotes} % provides \enquote for quotation marks

%% file: style/definitions.tex
\definecolor{blackberry}{HTML}{8D1D75}

\renewcommand{\vec}{\bm} % bold vector symbols
\DeclareSIUnit\parsec{pc}
\DeclareSIUnit\dex{dex}
\DeclareSIUnit\h{\mathnormal{h}}
\DeclareSIUnit\year{yr}
\DeclareSIUnit\years{yrs}
\DeclareSIUnit\arcsec{arcsec}
\DeclareSIUnit\arcmin{arcmin}
\DeclareSIUnit\Msun{M_\odot}
\DeclareSIUnit\Rsun{R_\odot}
\DeclareSIUnit\Lsun{L_\odot}
\DeclareSIUnit\Rvir{\mathnormal{R}_\mathrm{vir}}
\DeclareSIUnit\Rhalf{\mathnormal{R}_{1/2}}
\DeclareSIUnit\erg{erg}
\DeclareSIUnit\angstrom{\text{Å}}

\newcommand*{\Msun}{\ensuremath{\mathrm{M}_\odot}} % solar mass
\newcommand*{\Rsun}{\ensuremath{\mathrm{R}_\odot}} % solar radius
\newcommand*{\Lsun}{\ensuremath{\mathrm{L}_\odot}} % solar luminosity

\newcommand*{\Rhalf}{\ensuremath{R_{1/2}}} % half-mass radius